\documentclass [12pt]{iopart} 
\usepackage{iopams}
\usepackage{graphicx}
\begin{document}

\title{A Random Matrix Approach to Language Acquisition }
\author {A. Nicolaidis, Kosmas Kosmidis, Panos Argyrakis}
\address {Department of Physics, University of Thessaloniki, 54124 Thessaloniki, Greece}

\date{\today}

\begin{abstract}
 Since language is tied to cognition, we expect the linguistic structures to reflect patterns we encounter in nature and analyzed by physics. Within this realm we investigate the process of protolanguage acquisition, using analytical and tractable methods developed within physics. A protolanguage is a mapping between sounds and objects (or concepts) of the perceived world. This mapping is represented by a matrix and the linguistic interaction among individuals is described by a random matrix model. There are two essential parameters in our approach. The strength of the linguistic interaction $\beta$, which following Chomsky's tradition, we consider as a genetically determined ability, and the number $N$ of employed sounds (the lexicon size). Our model of linguistic interaction is analytically studied using methods of statistical physics and simulated by Monte Carlo techniques. The analysis reveals an intricate relationship between the innate propensity for language acquisition $\beta$ and the lexicon size $N$, $N \sim \exp(\beta)$. Thus a small increase of the genetically determined $\beta$ may lead to an incredible lexical explosion. Our approximate scheme offers an explanation for the biological affinity of different species and their simultaneous linguistic disparity.

\end{abstract}


\maketitle

\section {Introduction}
	Language has been a defining moment in the evolution of the human beings. It first appeared about 100000 years ago, in an eye-blink evolution, in the species Homo Sapiens. The sudden emergence and spread of language, like a viral epidemic, makes it hard to explain in terms of standard evolution, and echoes the reference to the evolution of language as the ``hardest problem of science'' \cite{Chomsky1965,Bickerton2007,nowak2000ebo,S.Fisher2006}. The language allowed an effective communication among the members of a human group, helped in transferring information from one generation to another, and even served as a systematic method to interpret the world, creating an endless semiotic process. The linguistic system is a highly generative system \cite{nowak2002cae}. Few phonemes form a large number of words. Words, following relatively few basic ``rules of composition'' (or a syntax), form an infinity of phrases and sentences. Thus, language enables us to transfer unlimited information. This limitlessness has been described as ``making infinite use of finite means'' \cite{Hauser2002,NowakKrak1999}.

Biology uses another exemplary generative system. Genomes consist of an alphabet of four nucleotides, which together with certain rules for how to produce proteins and organize cells, generates an unlimited variety of living organisms. Noam Chomsky, who revolutionized linguistic research, emphasized that the human faculty of language appears to be organized like the biological genetic code - hierarchical, generative, recursive, and virtually limitless with respect to its scope of expression. Our ability to understand and utter language is due to a universal grammar that is somehow hardwired within us \cite{Chomsky1975}. Language develops just like any other organ in the human body: an innate program, founded in a ``linguistic genotype'', supports linguistic growth, though the final ``linguistic phenotype'' is conditioned by experience. With these ideas in mind, one might wonder though, why our genetically closest relatives didn't develop something that is akin to language. Or, as it was  already put by Darwin in his `` Origin of Species''\cite{Darwin1859/1964}:

\textit{ ``not one author posed the question as to why in some animals the cognitive capabilities are developed more than in others, whereas such development should have been useful for all? Why monkeys did not acquire human intellectual capabilities?''}

In the present paper, we would like to draw attention to the oldest generative system, the physical world itself, and to its potential relevance for the language phenomenon. Despite its immense variety, nature can be analyzed and understood as a collection of few building blocks, the elementary particles (quarks, leptons, gauge particles). The elementary particles interact and form (or transformed to) larger compounds (nuclei, molecules, galaxies …) via the four well known interactions. We may view the elementary particles as constituting an ``alphabet'', and the interactions as providing the ``rules of composition'' (or ``grammatical rules'') to create the larger configurations. Within this analogy scheme, it is rather significant that the ancient Greeks were using the same word ($\sigma \tau o \iota \chi \varepsilon \iota \alpha$) to denote both the letters of the alphabet and the constitutive elements of the universe. Language cannot be separated from cognition, which reproduces the world. Linguistic devices expressing quantity, tense, comparison, temporal or logical relations, embody patterns encountered in nature. In an intense semiotic process, we constantly create mappings and analogies, sculpt outputs to match the external inputs. Nature then is reflected in our language and we dissect nature along lines laid down by language. This profound analogy, nature-human language \cite{A.Nicolaidis}, prompts us to use ideas and techniques encountered in physical theories, in order to study aspects of the linguistic dynamics.

As a first step in this approach, we consider the learning of a protolanguage, employing a dynamical scheme inspired by the random matrix approach and statistical physics. A protolanguage is a mapping between sounds and objects (or concepts) of the perceived world. A protolanguage may be represented by an association matrix and a population of individuals (humans or other animals) are using for their communication a specific association matrix \cite{Hurford1989,Plotkin2000}. Another individual (a newcomer, or a newborn) may use a different association matrix, selected randomly among the possible languages. We expect then that the interaction of the single individual with the population, to lead to a ``realignment'' of her (his) linguistic expression upon the language of the community. Our model simulates this process as a matrix-matrix interaction and the equilibrium reached is analyzed using the methods of statistical physics.

There is already a significant interdisciplinary research on the evolutionary aspects of human language. Such an interest is a direct consequence of the rapid advances in the field of complexity\cite{stauffer2006bsg}. Complex systems comprising of many interacting units are studied using the principles of Statistical Physics, even though the interacting units are no longer atoms as in traditional physics applications, but biological species\cite{droz2006pdh,penna1995maa}, human beings \cite{penna1995abs,gallos2005sos}, or financial markets \cite{mantegna1995sbi,mantegna1996taf}. Human language, which traditionally was viewed as a rather qualitative subject of study, fits adequately in the above dynamical framework. A study of the language, inspired by evolutionary dynamics, has been rigorously explored by Nowak and collegues\cite{nowak2000ebo,nowak2002cae}. The areas of study include also linguistic games \cite{baronchelli2008ida}, language competition between two \cite{patriarca2009iog,kosmidis2005lea,kosmidis2006lts,patriarca2004mlc} or more languages \cite{schulze2005mcs,schulze2006mcs,schwammle2005scl,schwammle2006pts,deoliveira2006bfl,deoliveira2006tmf,deoliveira2008csl,tuncay2007anm} to the quantification of language characteristics and their explanation from first principles \cite{cancho2003lea,kosmidis2006sma}.
The mathematical framework of language modeling and simulation has already given some rather intriguing results. Abrams and Strogatz \cite{abrams2003mdl} have proposed a simple model of non-linear differential equations which describe rather well the distribution of spoken languages and several extensions of this model have been subsequently studied \cite{patriarca2004mlc,stauffer2007mas}.
Several agent based models of language competition have been proposed \cite{schulze2005mcs,stauffer2006nea,deoliveira2006bfl} and the probability distribution of spoken languages has been described with considerable accuracy \cite{deoliveira2007bsa,tuncay2008tpo}. Recently, there has been an interesting attempt for a systematic study on the influence of the geography \cite{patriarca2009iog} on language competition, an original attempt to describe linguistic aspects in terms of random matrices \cite{tuncay2008tpo} and a study on the network properties of written human languages \cite{masucci2006npw}.
\\In section 2, we present in detail our model, including analytic approximations and Monte Carlo simulations. In section 3 we present the main results and discuss their importance. Our conclusions and directions for future work are presented in section 4.

\section{Model and Methods}
\subsection{Analytical description}
We imagine a group of individuals, which have established a simple communication system, by using sounds to encode meaning. Suppose that we have $N$ ``objects'' and each individual object is denoted by a distinct sound (a total of $N$ sounds). The mapping of objects to sounds is specified by an $N \times N$ active matrix $\bf{P}$, whose elements are either one or zero. For example, the entry $p_{ij}=1$ implies that the object $i$ is associated with the sound $j$.
Every time a speaker wishes to refer to object $i$ he is using the sound $j$.
Next to the mapping from object to sounds, there is another mapping from sounds back to objects, specified by the $N \times N$ passive matrix $\bf{Q}$. Again the elements $q_{nm}=1$ implies that a listener hearing sound $n$ will infer object $m$. Language involves both speaking and listening and the linguistic code of an individual $L(\bf{P},\bf{Q})$ is defined by these two matrices \cite{nowak2002cae,Hurford1989,Plotkin2000}. It is obvious that the maximum effectiveness of communication is achieved when the matrices $\bf{P}$ and $\bf{Q}$ are connected,
\begin{equation}
    p_{ij}=q_{ji}
\label{eq1}
\end{equation}
How many linguistic codes may we have? Matrix $\bf{P}$, as well as $\bf{Q}$, is constructed as a permutation matrix; that is, there is a single entry equal to one per row and column, all other entries being zero. There are $N!$ possible ways to associate $N$ objects to $N$ sounds and therefore $N!$ distinct linguistic codes. An established community advancing through sharing and exchanging information, is using the same language $L(\bf{P},\bf{Q})$. An individual, not a member of the community (a newcomer, or a newborn) might be using another language $L'(\bf{P'},\bf{Q'})$, chosen randomly among the $N!$ possibilities. Some of the associations object-sound (or sound-object) might be the same in both languages $L$ and $L'$, while others may be different.
\\The interaction between $A$ using language $L$ and $A'$ using language $L'$ is quantified \cite{nowak2002cae,Plotkin2000} by the ``communication energy'' $E$ 
\begin{equation}
    E(L,L')=-\frac{1}{2} \sum_{i,j} (p_{ij} q'_{ji}+ p'_{ij} q_{ji})
\label{eq2}
\end{equation}
$E$ is a direct measure of the communication success, the ability of $A$ to convey information to $A'$ and vice-versa. The first term $p_{ij} q'_{ji}$ denotes the possibility that speaker $A$ successfully communicates object $i$ to listener $A'$, while for the second term $p'_{ij} q_{ji}$ the speaker-listener relationship is reversed. If the same language is used, taking into account the condition Eq.(\ref{eq1}), we obtain 
\begin{equation}
    E(L,L)=-N
\label{eq3}
\end{equation}
marking the ideal communication. In general, for two different languages miscommunication occurs, resulting from the different assignments of objects to sounds. We expect in general that
 
\begin{equation}
   E(L,L') = -m ~~~ 0\leq m \leq N. 
\label{eq4}
\end{equation}
where $m$ is the number of common semantic associations the two linguistic codes have. It is expected that the single individual, in a continuous interaction with the surrounding environment which is using the definite code $L$, will increase the number of the common object-sound associations, thus stepping up the acquisition of the $L$ language. Within our model, this is achieved by providing a higher weight to the languages with an increased ``correct'' identification of objects to sounds. Following the experience from systems in equilibrium this statistical weight is chosen as $\exp(-\beta E)$. With $\beta$ we represent the strength of the linguistic interaction. Large values of $\beta$ favor the `` alignment'' of the linguistic choices, i.e. codes resembling $L$ are strongly favored. Low $\beta$ values allow the presence of a variety of languages. Our approach, considering the linguistic interaction as an intense one leading to equilibrium, suggests that we may use techniques from Statistical Physics. We define then the partition function as
\begin{equation}
    Z=\sum_{L'} \exp[-\beta E(L,L')]
\label{eq5}
\end{equation}
The summation is carried out over all possible `` linguistic states'' $L'$, while the code $L$ appears as a constant external field. Taking into account Eq.(\ref{eq4}), we obtain 
\begin{equation}
    Z=\sum_{m} g(m) \exp(\beta m)
\label{eq6}
\end{equation} 
where $g(m)$ is the multiplicity of languages sharing $m$ semantic associations with $L$. To evaluate $g(m)$, we start with the encoding in language $L$ considered as a permutation, and generate all other permutations keeping $m$ assignments fixed. We may select in
${N \choose m}$ different ways the $m$ fixed assignments, while for the other $N-m$ elements the permutation is a derangement. A derangement means that none of the elements may appear in its original position. The multiplicity $g(m)$ is then equal to 
\begin{equation}
g(m)=   
 \left( {\begin{array}{*{20}c}
   N  \\
   m  \\
\end{array}} \right) D(N-m)
\label{eq7}
\end{equation}
where the number of derangements is given by (the mathematical details may be found in the appendix):
\begin{equation}
    D(k)=k! (1-\frac{1}{1!}+\frac{1}{2!}-\frac{1}{3!}+\cdots+\frac{(-1)^k}{k!} )
\label{eq8}
\end{equation}
Notice that $D(0)=1, D(1)=0$, while for large $k$
\begin{equation}
    D(k)\simeq \frac{k!}{e}
\label{eq9}
\end{equation}
An accurate expression for the partition function is obtained then

\begin{equation}
    Z \simeq \frac{N!}{e} \sum_{m=0}^{N-2} \frac{e^{m \beta}}{m!} +\exp(\beta N)
\label{eq10}
\end{equation}
All measurable quantities concerning our system can be derived from the partition function. 
Language acquisition can be measured by studying the average number of common associations  $\langle m \rangle$, given by
\begin{equation}
    \langle m \rangle= \frac{ \sum m g(m) e^{\beta m}}{\sum g(m) e^{\beta m}}=\frac{1}{Z} \frac{\partial Z}{\partial \beta}
\label{eq11}
\end{equation}
Fluctuations around the mean value can be estimated as
\begin{equation}
 \Delta m^2 =\langle m^2 \rangle -\langle m \rangle^2 =\frac{\partial^2 \ln Z}{\partial \beta^2}
\label{eq12}
\end{equation}
Our simple linguistic system reveals interesting correlations between the interaction 
strength $\beta$ and the size of the lexicon $N$. Consider first the case of small $\beta$, where  many terms contribute in eq.(\ref{eq10}). The terms entering in the summation build up an exponential which dominates, and the partition function is then approximated  by

\begin{equation}
    Z \simeq \frac{N!}{e} \exp  [\exp(\beta)] \ \ \ \ \ \ \mbox{(small $\beta$)}
\label{eq13}
\end{equation}
From Eq. (\ref{eq11}) we find
\begin{equation}
    \langle m \rangle \simeq \exp(\beta) \ \ \ \ \mbox{(small $\beta$)}
\label{eq14}
\end{equation}
We notice that the number of acquired words increases exponentially with the interaction strength $\beta$. Another way of stating our result is that a small increase in $\beta$, which may be considered as an innate propensity for language acquisition, provokes an exponential growth of the size of the available lexicon. The spread around the average value $\langle m \rangle$ is, using Eq. (\ref{eq12})-(\ref{eq13}),

\begin{equation}
    \Delta m^2 \simeq \exp(\beta) \ \ \ \ \mbox{(small $\beta$)}
\label{eq15}
\end{equation}

The spread is significant since many linguistic states contribute to the mean value. 
\\For large $\beta$ values the important contributions are coming from languages very similar to $L$.
In this case
\begin{equation}
    Z \simeq e^{\beta N} + \frac{N(N-1)}{2} e^{\beta (N-2)} \ \ \ \ \mbox{(large $\beta$)}
\label{eq16}
\end{equation}
The mean value  $\langle m \rangle$ is
\begin{equation}
    \langle m \rangle \simeq N \ \ \ \ \mbox{(large $\beta$)}
\label{eq17}
\end{equation}
while the spread around the mean value decays exponentially as $\beta$ increases. The crossover between the two regimes, small $\beta$ vs large $\beta$ values, occurs at
\begin{equation}
    \beta_{cr} \simeq \ln N
\label{eq18}
\end{equation}
The underlying dynamics is manifested when we consider the entropy $S$
\begin{equation}
    S=-\beta \frac{\partial \ln Z}{\partial \beta}  + \ln Z
\label{eq19}
\end{equation}

Using Eq. (\ref{eq13}) we find
\begin{equation}
    S=N \ln N - \beta \exp(\beta)
\label{eq20}
\end{equation}

At small $\beta$ values the entropy is large, reflecting that all $N!$ codes contribute, while as $\beta$ approaches $\beta_{cr}$ the entropy becomes zero since only one code contributes.
The theoretical analysis is supported by a detailed Monte-Carlo simulation.

\subsection{Monte Carlo Simulations} 
We simulate the learning process described by the above model using the following algorithm. First an initial sequence of integers from $1$ to $N$ is chosen at random to represent the ``optimal'' language which has to be learned (understood and memorized) by a learning agent. Then, another permutation is chosen to represent the language of this learning agent. When the two sequences have the same number at the same position this is considered to be a success, meaning that the learner has correctly identified the meaning of this word and associated it with the proper object. We compare the two sequences and count the number $m$ of successes. Then, a random pair of the $N$ elements of the learners vocabulary is chosen and their position is interchanged. Again the new number of successes  $m_{new}$ is calculated. If $m_{new}$ is greater than the previous $m_{old}$ the flip is accepted with probability one. Otherwise it is accepted with probability 
\begin{equation}
 p=\exp( \beta \Delta m ) 
\label{eq21}  
\end{equation}
where $\Delta m=(m_{new}-m_{old})$, as is typically done in Statistical Mechanics simulations with Metropolis dynamics.
We continue this iterative process until the system reaches an equilibrium state and we calculate 
$\langle m \rangle$ and Var($m$) by averaging our results over 500 initial system realizations.

\section{Results and Discussion}
\begin{figure}
\begin{center}
\includegraphics[height=8.5 cm,width=8.5cm]{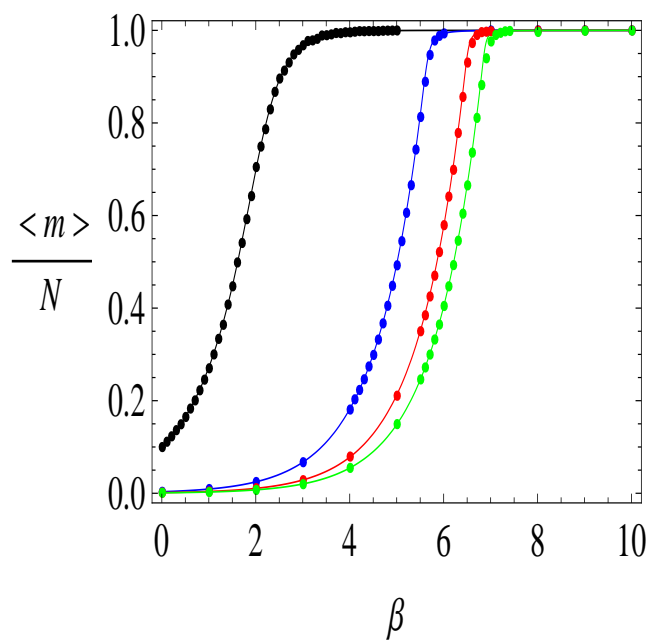}
\end{center}
\caption{Mean number of words (normalized), $\langle m \rangle /N$ vs parameter $\beta$ for $N=10,300,700,1000$ (black, blue, red, green). Symbols are results of Monte Carlo Simulation and solid lines are numerical solutions of Eq.(\ref{eq11}) }
\label{fig1}
\end{figure}

In Fig. \ref{fig1}, we plot the mean number of words $\langle m \rangle$ after the system has reached the equilibrium state divided by the language size $N$ for several system sizes, namely for $N=10, 300, 700, 1000$. The analytic results (solid lines) and the Monte Carlo Simulations (points) are in excellent agreement. We observe that for small $\beta$, the fraction  $\frac{\langle m \rangle}{N}$ tends to zero while for large $\beta$ it becomes equal to one and that there exists a crossover between the two states at a crossover value $\beta_{cr}$ which depends on $N$. In fact, it seems that $\beta_{cr}$ increases monotonically with increasing $N$ and that for given $\beta$ there is always a language size $N(\beta)$, such as below $\beta$ the learning fraction is in the ``zero'' state and above it is in the ``one'' state. This aspect is a characteristic of a crossover phenomenon in contrast to a phase transition where there is a critical value of a control parameter which does not depend on system size in such a manner and which remains finite even for infinitely large system sizes.      

\begin{figure}
\begin{center}
\includegraphics[height=8.5 cm,width=8.5cm]{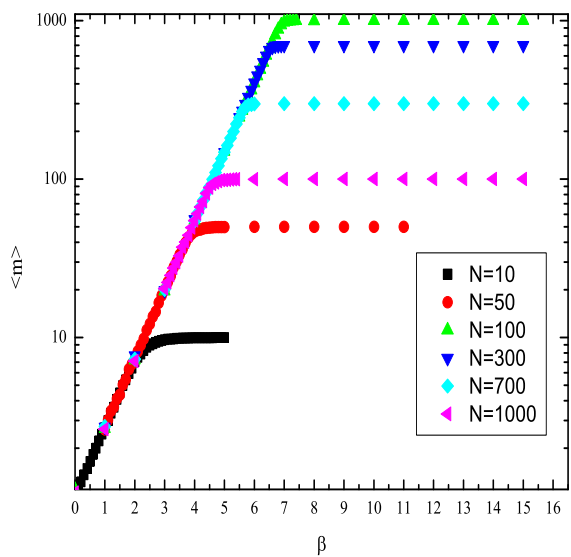}
\end{center}
\caption{Log-Linear plot of the mean number of words $\langle m \rangle$ vs $\beta$ for $N=10, 50, 100, 300, 700, 1000$. Notice the data collapse for small $\beta$ values indicative of an exponential scaling independent of $N$.  }
\label{fig2}
\end{figure}

In order to check the validity of our approximation for small language sizes we present, in Fig. \ref{fig2}, a log-linear plot of the mean number of words $\langle m \rangle$ in equilibrium as a function of $\beta$ for system sizes $N=10, 50, 100, 300, 700, 1000$. We observe that for small $\beta$ 
the points are in straight lines indicating an exponential dependence of $\langle m \rangle$ on $\beta$. Moreover, the data collapse indicates the independence of $\langle m \rangle$ on the language size $N$ in complete agreement to our analytical predictions, Eq. (\ref{eq14}).   

\begin{figure}
\begin{center}
\includegraphics[height=8.5 cm,width=8.5cm]{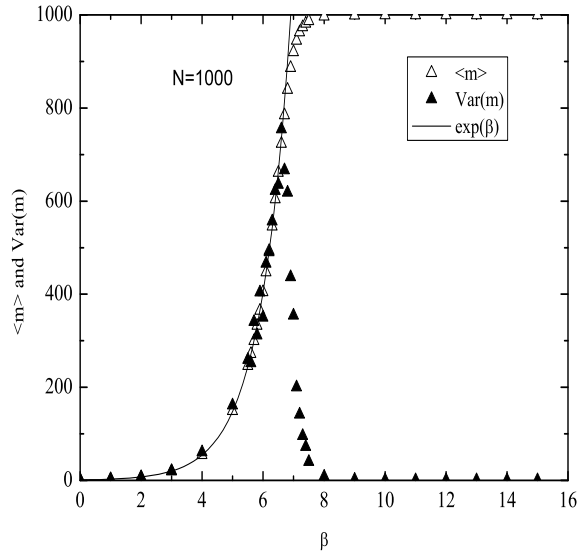}
\end{center}
\caption{Mean number of words $\langle m \rangle$ and variance Var$(m)$ vs $\beta$ for language size $N=1000$. Triangles are Monte Carlo Simulation results for $\langle m \rangle$ (white) and Var$m)$ (black) and the solid line is equal to $\exp(\beta)$. }
\label{fig3}
\end{figure}

Next, we study the variance Var$(m)$ of the vocabulary size $m$. Figure \ref{fig3} shows the number of words $\langle m \rangle$ and variance Var$(m)$ vs $\beta$ for language size $N=1000$.  The triangles are Monte Carlo simulation results and the solid line is equal to $\exp(\beta)$. The collapse of the points indicates that up to a characteristic crossover value $\beta_{cr}$ both $\langle m \rangle$ and Var$(m)$ increase with increasing $\beta$ and they are both equal to $\exp(\beta)$ in agreement with our analytical prediction. Above $\beta_{cr}$ the equilibrium vocabulary size assumes with high probability its maximum value, thus there is a decrease in the fluctuations of $m$ while a sharp maximum of Var$(m)$ is observed at $\beta_{cr}$.   

\begin{figure}
\begin{center}
\includegraphics[height=8.5 cm,width=8.5cm]{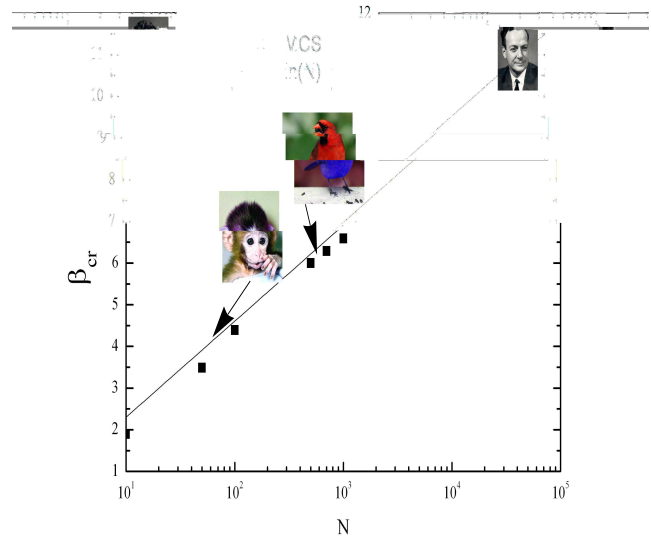}
\end{center}
\caption{Crossover value $\beta_{cr}$ versus the language size $N$. Squares are estimates of the crossover from the maximum of the Var$(m)$ and the solid line is equal to $\ln N$. Human average vocabulary is $\approx 60000$ words, while birds use roughly $\approx 1000$ sounds and apes understand even less than that.  }
\label{fig4}
\end{figure}

Finally, we examine how the crossover value $\beta_{cr}$ scales with the language size $N$. We determine $\beta_{cr}$ from the position of the maximum of Var$(m)$. Figure \ref{fig4} shows that $\beta_{cr} \sim \ln N$ in quite good agreement with the analytical prediction. The physical significance of $\beta_{cr}$ is that it determines a minimum of linguistic ability that is required by an individual for efficiently learning a language of size $N$. This scaling implies that a small increase of the ability parameter $\beta$ will have a profound impact on language learning as it may lead from a ``zero'' state for the effective vocabulary (below $\beta_{cr}$) to the ``one'' state of successful learning (above $\beta_{cr}$). 
\\Our model allows a comparative analysis of animal communication.
\\Following Chomsky \cite{Chomsky1965,Chomsky1975}, a strong connection between biology and linguistics has been promoted, with genetically determined rules controlling the linguistic ability. The parameter $\beta$ represents in an effective way this genetically determined linguistic ability and different species have different â values. A given species, qualified by linguistic ability $\beta$, may acquire and use a language consisting of up to N words, where
\begin{equation}
    N \sim \exp(\beta)
\label{eq22}
\end{equation}                                                                                  
Notice that a small increase in $\beta$, the biological ability for language acquisition, induces an exponential growth of the size of the available lexicon. Trained apes can learn 50-200 words, the most well known case being the bonobo chimpanzee named Kinzi \cite{Savage-Rumbaugh1998}. This size of the lexicon is reproduced by a $\beta$ of about 4. Songbirds display a richer lexicon of about 700 words \cite{Gentner2006}, corresponding to a value 6 for $\beta$. An average high-school graduate has a lexicon of about 60000 words \cite{nowak2000ebo,Hauser2002}, giving a $\beta$ value close to 11 for the human species.  We observe that a relatively small range in the genetically determined $\beta$ parameter gives rise to immense variations in the size of the employed lexicon(see fig. \ref{fig4}). Thus we have an approximate scheme, which can accommodate the biological affinity of some species and their linguistic disparity.

\section{Conclusions and future directions}
We are dealing with language and it is not appropriate to consider it as an isolated system. Rather we hope to capture aspects of the complex linguistic phenomenon by resorting to a highly interdisciplinary method. In our paper we suggested that models and techniques developed within physics might be useful in deciphering the language riddle. The rationale behind the indicated course is that since language is strongly tied to cognition, we expect the linguistic structures to reflect structures and patterns we encounter in nature and analyzed by physics. This profound interrelationship nature - human language is a permanent and continuous one and lies at the very foundation of the ``intelligibility'' of the universe. 
As a first step we considered the most simple language, a protolanguage, which is essentially a mapping between sounds and objects. This mapping is represented by a matrix and the language interaction is simulated by random matrix mechanics. The suggested interaction Hamiltonian between the matrices is (see eq. \ref{eq2})
\begin{equation}
 H=(1/2) Tr (\mathbf{PQ'} + \mathbf{P' Q}) 
\label{eq23}  
\end{equation}

Our simple model bears great resemblance to a well known and extensively studied problem in physics, magnet-magnet interaction. A magnet may have one direction in space, chosen among a given set of possible directions. When many magnets are brought together, it is expected that the interaction among the magnets to lead the magnets to acquire a common direction in space, rather than each magnet having its own direction. This common field is described as mean field and an individual field (a magnet, or a particle) interacts with this average mean field. A particular matrix version of the mean field technique may be found in ref.\cite{Brezin1980} , and our model Hamiltonian is very similar to theirs. In a similar vein, a protolanguage appears as a specific choice among a huge number ($N!$) of possibilities. Social interaction among the different partners, each using its own protolanguage, will lead eventually to the adoption of a unique collective ``mean protolanguage'', $L(P,Q)$ in our case. It is with this ``mean protolanguage'' that an individual will interact, the interaction being described by eq. \ref{eq23}. Random matrices have been widely used in Nuclear and Particle Physics and in general in systems involving large numbers of degrees of freedom \cite{Wigner1967,Mehta2004}. Matrix models are directly linked to string theory \cite{Bigatti}, the theory unifying all interactions in nature \cite{Green1986}. Also it has been shown recently that relational logic and category theory are expressed by matrix models \cite{Nicolaidis2009}. Thus, our proposal opens the possibility for a fruitful interaction between linguistics and advanced sectors of theoretical and mathematical physics.

We adopted Chomsky's vision that language acquisition is rooted in innate structures and innateness comes in degrees. This linguistic innateness is represented in our model by the effective parameter $\beta$, having different values for the different biological species. We can only advance hypotheses about what lies behind the dispersion of $\beta$ values, the innate propensity for language acquisition. It has been suggested that the human brain, being relatively larger than that of other primates, runs a significantly larger number of neural interconnections \cite{Darwin1871/1981}, leading to a high $\beta$ value for the humans. Along a different line, neurobiologists have identified the gene FOXP2 as directly affecting the language ability in humans \cite{Bickerton2007,S.Fisher2006,Vargha-Khadem2005}. The presence and the specific functioning of similar genes in other primates and the songbirds is of prime importance \cite{Vargha-Khadem2005,Miller2008}. Bipedalism also has been considered as a factor favoring the development of language. Upright posture sets the hands free for alternate uses \cite{Macneilage1998}and provides a frontal and wide view of the environment, thus increasing the stimulus for cognition and symbolic expressions. 

Spoken languages leave no fossils and consequently it is not easy to infer the language evolution. But as Simon Conway Morris argues, ``it would be strange if my fingers and eyes were to have an evolutionary origin but not my capacity to speak''\cite{Conway}. Two evolutionary scenarios have been advanced, a gradual and mosaic one, where language follows the pattern of most evolutionary events (like the long evolution of eye) and an abrupt one, where language emerges in a single step process \cite{Bickerton2007}. Our work offers a further step of this intricate issue. The relationship between the innate propensity for language ($\beta$) and the lexicon richness  is a continuous one, as displayed in fig. \ref{fig1}. One notices, though, the abrupt transition from poor linguistic achievement ($\beta < \beta_{cr}$) to high linguistic achievement ($\beta > \beta_{cr}$). A small increase of $\beta$ may lead to an incredible evolutionary leap, which may be qualified as a ``lexical big bang''. In that way we may interpret the apparent language discontinuity between humans and the other hominoids. 

More than 50 years of research using classical training studies demonstrates that animals (apes, parrots, pigeons, rats) can acquire a number of words or concepts\cite{Hauser2002}. With regard to number quantification animals can represent numbers up to a maximum (around 9)\cite{Gallistel1990}. As the target number increases, the standard deviation around the matched mean increases accordingly. This spread around the mean value is reproduced by our model, see eqs. \ref{eq15} and \ref{eq14} for small $\beta$. On the other hand, humans are unique in the ability to show an open-ended quantification skill, including discrete infinity among the numbers. We attribute again this human capacity to a corresponding large $\beta$ value.

There is a strong tendency to advocate a modular dissociation between lexicon and grammar, between protolanguage and fully developed language. Bates and Goodman have provided evidence that the emergence of grammar is highly dependent upon the lexicon size\cite{Bates1997}. Thus the degree of grammatical competence acquired by children is strictly linked to the lexical stage at which they are. Children with lexicons under 300 words have very restricted grammatical abilities. Viewed in this light, chimpanzees, with a lexicon of 200 words, appear to be arrested at a point in lexical development when grammar is still at a very simple level\cite{Bates1997}. This type of approach is corroborated by the experimental finding that songbirds, possessors of a richer lexicon composed of 700 sounds, recognize acoustic patterns defined by a recursive, self-embedding, context-free grammar\cite{Gentner2006}. Further along is the language of the human primate, with a much larger lexicon and considerably richer grammar. A comparison reveals that while on biological grounds we are close to the other primates, on linguistic grounds we are closer to birds (the human as a singing ape was described by Darwin back in 1871\cite{Darwin1871/1981}). Fig \ref{fig4}, displaying the genetic propensity for language $\beta$ vs lexicon size $N$, may be viewed with the coordinate $N$ representing also the grammatical complexity.

Our exploration of reality is always mediated by language or a general semiotic process. Next to the real world, we create an entire world of symbols, organized internally by the different forms of language. The symbolic world is substantiated by individual cognitive units (neurons), joined and operated by vastly unknown physical mechanisms. Or as Noam Chomsky put it: ``We know very little about what happens when $10^{10}$ neurons are crammed into something of the size of a basketball, with further conditions imposed by the specific manner in which this system developed over time\cite{Chomsky1975}''.
And later:
``It may be that at some remote period a mutation took place that gave rise to the property of discrete infinity, to be explained in terms of the property of physical mechanisms, now unknown\cite{Chomsky1987}''.
Symbols and words are organized into finite strings (sentences), following a finite number of grammatical rules, through the recursive application of these rules. We consider that the grammatical parsing of languages bears resemblance to the parsing of the natural processes occurring in the world. Both of them may be simulated by random matrix dynamics, involving interaction terms more complex than the one considered in the present paper (eq. \ref{eq23}). We hope that this type of approach, incorporating ideas and models from physics into the language research, will appear
fruitful and interesting in the future.

\appendix
\section*{Appendix}

A derangement is a permutation in which none of the elements of the set appear in their original positions. Or considered as a bijection $f: S \rightarrow S$ , the derangement does not allow an element $x \in S $ with $f(x)=x$. To find the number of derangements of an $n$-element set $S$, 
the inclusion- exclusion principle has been used. The set of all permutations $P$ of the set $S$ has cardinality $|P|=n!$. To obtain the number of derangements we have to subtract from the total number of permutations those which map an element to itself. Let us call $A_i (1 \leq i \leq n) $ the set of all permutations that map the $i$th element to itself. Then $\sum |A_i|= {n \choose 1} (n-1)!$. This process leads to underestimation since the subtraction involves twice  the permutations having two fixed points. We should add then 
$ \sum_{i < j} |A_i  \bigcap A_j|={n \choose 2}(n-2)!$.

Again, we reach an underestimation, since in the previous summation we have included twice the permutations involving three fixed points. This type of analysis continues until we reach the $n^{th}$ term, and the number of derangements emerge as a sum with alternating signs 
\begin{eqnarray}
D(n)& = & n!-{n \choose 1}(n-1)!+{n \choose 2}(n-2)!- \cdots + (-1)^n {n \choose n}(n-n)! \nonumber \\
& = &  n! (\frac{1}{0!} -\frac{1}{1!} + \frac{1}{2!}- \cdots +\frac{(-1)^n}{n!})\nonumber
\end{eqnarray}

Notice also that
$$ \sum_{m=0}^{n} {n \choose m} D(n-m)=n! $$

\section*{References}

\bibliography{Language}

\begin{thebibliography}{10}

\bibitem{Chomsky1965}
Noam Chomsky.
\newblock {\em Aspects of the theory of syntax}.
\newblock The MIT Press, Cambridge, MA, 1965.

\bibitem{Bickerton2007}
D.~Bickerton.
\newblock Language evolution; a brief guide for linguists.
\newblock {\em Lingua}, 117:510--526, 2007.

\bibitem{nowak2000ebo}
MA~Nowak.
\newblock Evolutionary biology of language.
\newblock {\em Phil. Trans. of the Royal Soc. of London Series B-Biol. Sci.},
  355(1403):1615--1622, NOV 29 2000.

\bibitem{S.Fisher2006}
S.~Fisher and G.~Marcus.
\newblock The eloquent ape: genes, brains and the evolution of language.
\newblock {\em Nature Reviews}, 7:9--20, 2006.

\bibitem{nowak2002cae}
MA~Nowak, NL~Komarova, and P~Niyogi.
\newblock Computational and evolutionary aspects of language.
\newblock {\em Nature}, 417(6889):611--617, JUN 6 2002.

\bibitem{Hauser2002}
M.~Hauser, N.~Chomsky, and T.~Fitch.
\newblock The faculty of language: what is it, who has it, and how did it
  evolve?
\newblock {\em Science}, 298:1569--1579, 2002.

\bibitem{NowakKrak1999}
Martin~A. Nowak and David~C. Krakauer.
\newblock The evolution of language.
\newblock {\em Proc. Natl. Acad. Sci.}, 96(14):8028--8033, July 1999.

\bibitem{Chomsky1975}
Noam Chomsky.
\newblock {\em Reflections of language}.
\newblock Pantheon Books, New York, 1975.

\bibitem{Darwin1859/1964}
Charles Darwin.
\newblock {\em On the origin of species}.
\newblock Harvard University press, Cambridge, MA, facsmile edition edition,
  1859/1964.

\bibitem{A.Nicolaidis}
A.~Nicolaidis.
\newblock {\em The Trinity and an Entangled World: Relationality in Physical
  Science and Theology}, chapter Relational Nature.
\newblock Grand Rapids, MI, Eerdmans, (forthcoming).

\bibitem{Hurford1989}
J.~Hurford.
\newblock {\em Lingua}, 77:187--222, 1989.

\bibitem{Plotkin2000}
Joshua~B. Plotkin and Martin~A. Nowak.
\newblock Language evolution and information theory.
\newblock {\em Journal of Theoretical Biology}, 205(1):147--159, July 2000.

\bibitem{stauffer2006bsg}
D.~Stauffer and S.M. de~Oliveira.
\newblock {\em Biology, Sociology, Geology by Computational Physicists}.
\newblock Elsevier, 2006.

\bibitem{droz2006pdh}
M.~Droz and A.~Pe{\c{}}kalski.
\newblock Population dynamics in heterogeneous conditions.
\newblock {\em Physica A: Statistical Mechanics and its Applications},
  362(2):504--512, 2006.

\bibitem{penna1995maa}
TJP Penna, SM~de~Oliveira, and D~Stauffer.
\newblock Mutation accumulation and the catastrophic senescence of the pacific
  salmon.
\newblock {\em Physical Review E}, 52(4, Part A):R3309--R3312, OCT 1995.

\bibitem{penna1995abs}
TJP Penna.
\newblock A bit-string model for biological aging.
\newblock {\em Journal of Statistical Physics}, 78(5-6):1629--1633, MAR 1995.

\bibitem{gallos2005sos}
LK~Gallos.
\newblock Self-organizing social hierarchies on scale-free networks.
\newblock {\em International Journal of Modern Physicsc C}, 16(8):1329--1336,
  AUG 2005.

\bibitem{mantegna1995sbi}
RN~Mantegna and HE~Stanley.
\newblock Scaling behavior in the dynamics of an economic index.
\newblock {\em Nature}, 376(6535):46--49, JUL 6 1995.

\bibitem{mantegna1996taf}
RN~Mantegna and HE~Stanley.
\newblock Turbulence and financial markets.
\newblock {\em Nature}, 383(6601):587--588, OCT 17 1996.

\bibitem{baronchelli2008ida}
Andrea Baronchelli and Vittorio Loreto.
\newblock In-depth analysis of the naming game dynamics: The homogeneous mixing
  case.
\newblock {\em International Journal of Modern Physicsc C}, 19(5):785--812, MAY
  2008.

\bibitem{patriarca2009iog}
Marco Patriarca and Els Heinsalu.
\newblock Influence of geography on language competition.
\newblock {\em Physica A: Statistical Mechanics and its Applications},
  388(2-3):174--186, JAN 15 2009.

\bibitem{kosmidis2005lea}
K.~Kosmidis, J.M. Halley, and P.~Argyrakis.
\newblock Language evolution and population dynamics in a system of two
  interacting species.
\newblock {\em Physica A: Statistical Mechanics and its Applications},
  353:595--612, 2005.

\bibitem{kosmidis2006lts}
K.~Kosmidis, A.~Kalampokis, and P.~Argyrakis.
\newblock Language time series analysis.
\newblock {\em Physica A: Statistical Mechanics and its Applications},
  370(2):808--816, 2006.

\bibitem{patriarca2004mlc}
M~Patriarca and T~Leppanen.
\newblock Modeling language competition.
\newblock {\em Physica A: Statistical Mechanics and its Applications},
  338(1-2):296--299, JUL 1 2004.

\bibitem{schulze2005mcs}
C~Schulze and D~Stauffer.
\newblock Monte carlo simulation of the rise and the fall of languages.
\newblock {\em International Journal of Modern Physicsc C}, 16(5):781--787, MAY
  2005.

\bibitem{schulze2006mcs}
Christian Schulze and Dietrich Stauffer.
\newblock Monte carlo simulation of survival for minority languages.
\newblock {\em Advances in Complex Systems}, 9(3):183--191, SEP 2006.

\bibitem{schwammle2005scl}
V.~Schw{\"a}mmle.
\newblock Simulation for competition of languages with an aging sexual
  population.
\newblock {\em International Journal of Modern Physics C}, 16(10):1519--1526,
  2005.

\bibitem{schwammle2006pts}
V.~Schw{\"a}mmle.
\newblock Phase transition in a sexual age-structured model of learning foreign
  languages.
\newblock {\em International Journal of Modern Physicsc C}, 17(01):103--111,
  2006.

\bibitem{deoliveira2006bfl}
VM~de~Oliveira, PRA Campos, MAF Gomes, and IR~Tsang.
\newblock Bounded fitness landscapes and the evolution of the linguistic
  diversity.
\newblock {\em Physica A: Statistical Mechanics and its Applications},
  368(1):257--261, AUG 1 2006.

\bibitem{deoliveira2006tmf}
VM~de~Oliveira, MAF Gomes, and IR~Tsang.
\newblock Theoretical model for the evolution of the linguistic diversity.
\newblock {\em Physica A: Statistical Mechanics and its Applications},
  361(1):361--370, FEB 15 2006.

\bibitem{deoliveira2008csl}
Paulo~Murilo Castro De~Oliveira, Dietrich Stauffer, Soren Wichmann, and
  Suzana~Moss De~Oliveira.
\newblock A computer simulation of language families.
\newblock {\em Journal of Linguistics}, 44(3):659--675, NOV 2008.

\bibitem{tuncay2007anm}
Caglar Tuncay.
\newblock A new model for competition between many languages.
\newblock {\em International Journal of Modern Physicsc C}, 18(7):1203--1208,
  JUL 2007.

\bibitem{cancho2003lea}
R.F. Cancho and R.V. Sole.
\newblock Least effort and the origins of scaling in human language.
\newblock {\em Proceedings of the National Academy of Sciences},
  100(3):788--791, 2003.

\bibitem{kosmidis2006sma}
K.~Kosmidis, A.~Kalampokis, and P.~Argyrakis.
\newblock Statistical mechanical approach to human language.
\newblock {\em Physica A: Statistical Mechanics and its Applications},
  366:495--502, 2006.

\bibitem{abrams2003mdl}
D.M. Abrams and S.H. Strogatz.
\newblock Modelling the dynamics of language death.
\newblock {\em Nature}, 424(2):900, 2003.

\bibitem{stauffer2007mas}
Dietrich Stauffer, Xavier Castello, Victor~M. Eguiluz, and Maxi~San Miguel.
\newblock Microscopic abrams-strogatz model of language competition.
\newblock {\em Physica A: Statistical Mechanics and its Applications},
  374(2):835--842, FEB 1 2007.

\bibitem{stauffer2006nea}
D.~Stauffer, C.~Schulze, F.~W.~S. Lima, S.~Wichmann, and S.~Solomon.
\newblock Non-equilibrium and irreversible simulation of competition among
  languages.
\newblock {\em Physica A: Statistical Mechanics and its Applications},
  371(2):719--724, NOV 15 2006.

\bibitem{deoliveira2007bsa}
P.~M.~C. de~Oliveira, D.~Stauffer, F.~W.~S. Lima, A.~O. Sousa, C.~Schulze, and
  S.~Moss de~Oliveira.
\newblock Bit-strings and other modifications of viviane model for language
  competition.
\newblock {\em Physica A: Statistical Mechanics and its Applications},
  376:609--616, MAR 15 2007.

\bibitem{tuncay2008tpo}
Caglar Tuncay.
\newblock The physics of randomness and regularities for languages in terms of
  random matrices.
\newblock {\em EPL}, 82(2), APR 2008.

\bibitem{masucci2006npw}
AP~Masucci and GJ~Rodgers.
\newblock Network properties of written human language.
\newblock {\em Physical Review E}, 74(2):26102, 2006.

\bibitem{Savage-Rumbaugh1998}
S.~Savage-Rumbaugh, S.~Shanker, and T.~Taylor.
\newblock {\em Apes, language and the human mind}.
\newblock Oxford University Press, 1998.

\bibitem{Gentner2006}
Timothy~Q. Gentner, Kimberly~M. Fenn, Daniel Margoliash, and Howard~C. Nusbaum.
\newblock Recursive syntactic pattern learning by songbirds.
\newblock {\em Nature}, 440:1204--1207, 2006.

\bibitem{Brezin1980}
E.~Brezin and D.~Gross.
\newblock The external field problem in the large n limit of qcd.
\newblock {\em Phys. Lett. B}, 97:120--124, 1980.

\bibitem{Wigner1967}
E.~Wigner.
\newblock Random matrices in physics.
\newblock {\em SIAM Review}, 9:1--23, 1967.

\bibitem{Mehta2004}
M.~Mehta.
\newblock {\em Random Matrices}.
\newblock Academic, 2004.

\bibitem{Bigatti}
D.~Bigatti and L.~Susskind.
\newblock Review of matrix theory.
\newblock {\em arxiv:hep-th/9712072}.

\bibitem{Green1986}
M.~Green, J.~Schwarz, and E.~Witten.
\newblock {\em Superstring Theory}.
\newblock Cambridge University Press, Cambridge, 1986.

\bibitem{Nicolaidis2009}
A.~Nicolaidis.
\newblock Categorical foundation of quantum mechanics and string theory.
\newblock {\em Int. J. Mod. Phys. A}, 24:1175--1183, 2009.

\bibitem{Darwin1871/1981}
Charles Darwin.
\newblock {\em The descent of man and selection in relation to sex}.
\newblock Princeton University Press, 1871/1981.

\bibitem{Vargha-Khadem2005}
F.~Vargha-Khadem, D.~Gadian, A.~Copp, and M.~Mishkin.
\newblock Foxp2 and the neuroanatomy of speech and language.
\newblock {\em Nature Reviews}, 6:131--138, 2005.

\bibitem{Miller2008}
Jane Miller, Elizabeth Spiter, Michael~Christopher Condro, Ryan~T.
  Dosumu-Johnson, Daniel~H. Geschwind, and Stephanie~Ann White.
\newblock Birdsong decreases protein levels of foxp2, a molecule required for
  human speech.
\newblock {\em Journal of Neurophysiology}, 100:2015--2025, 2008.

\bibitem{Macneilage1998}
P.~Macneilage.
\newblock {\em Approaches to the evolution of language}.
\newblock Cambridge University Press, Cambridge, 1998.

\bibitem{Conway}
S.~Conway Morris.
\newblock Does evolution explain human nature?
\newblock online at www.templeton.org/evolution.

\bibitem{Gallistel1990}
C.~Gallistel.
\newblock {\em The organization of learning}.
\newblock MIT Press, Cambridge, MA, 1990.

\bibitem{Bates1997}
E.~Bates and J.~Goodman.
\newblock On the inseparability of grammar and the lexicon: evidence from
  acquisition, aphasia and real-time processing.
\newblock {\em Language and Congitive Processes}, 12:507--584, 1997.

\bibitem{Chomsky1987}
Noam Chomsky.
\newblock {\em Language and problems of knowledge: the Managua
  lectures.(Current studies in linguistics).}, volume~16.
\newblock MIT Press, Cambridge, MA, 1987.

\end{thebibliography}

\end{document}